\documentclass[pra, reprint, showkeys, amsmath,amssymb, aps,]{revtex4-2}
\usepackage{graphicx}
\graphicspath{{./media/1_Geometry/},{./media/2_particle_shape/},{./media/4_packing_density/Sphere/}{./media/4_packing_density/}}
\usepackage{subcaption}

\usepackage[colorlinks=true, linkcolor=blue, citecolor=blue, urlcolor=blue]{hyperref} 
\usepackage{placeins}

\begin{document}
\title{Impact of friction and grain shape on the morphology of sheared granular media}

\author{Huzaif Rahim}
  \email{huzaif.rahim@fau.de}
\author{Sudeshna Roy}
\author{Thorsten P\"oschel}

\affiliation{
Institute for Multiscale Simulation, \\ Friedrich-Alexander-Universit\"at Erlangen-N\"urnberg, \\ Cauerstrasse 3, 91058 Erlangen, Germany
}

\begin{abstract} 
The interplay between dilatancy and particle alignment in sheared granular materials composed of non-spherical particles leads to morphological inhomogeneity. Dilatancy, driven by interparticle friction, causes the packing to expand, while particle alignment tends to densify it. We examine the influence of friction, particle aspect ratio, and initial packing conditions on the steady-state particle alignment and packing density. Unlike spherical particles, non-spherical particles with higher AR exhibit either dilatancy or compaction under shear, leading to spontaneous heaping or depression formation. We analyzed the evolution of packing density to identify whether dilatancy or compression prevails within the shear band.
\end{abstract}
\date{\today}
\maketitle

\section{Introduction}
Granular materials composed of non-spherical particles display complex packing and flow behaviors shaped by particle geometry, size, and frictional interactions \cite{kou2017granular, jaeger1996granular, duran2012sands, anthony2005influence, yu1998prediction, cheng2000dynamic,krishnaraj2016dilation}. High friction inhibits particle motion, leading to less dense packings, while low friction facilitates particle rearrangement, resulting in denser packings \cite{tory1968anisotropy, visscher1972random, tory1973simulated, adams1972computation}. For example, non-frictional spheres with friction coefficient $\mu = 0$ reach a packing density of approximately 0.64, which decreases to around 0.55 at $\mu = 0.5$ \cite{silbert2002geometry, zhang2001simulation, jerkins2008onset, jin2010first, farrell2010loose, silbert2010jamming}. Under gravity, non-spherical particles, such as elongated grains (sticks), typically pack less densely than spheres \cite{kou2017granular, zhang2001simulation, silbert2002geometry, salerno2018effect}. When subjected to tapping, shearing, or vibration, sticks can align and rearrange, resulting in increased packing density \cite{nagy2023flow, pol2022kinematics, borzsonyi2012orientational, borzsonyi2013granular, campbell2011elastic, guo2013granular}. This alignment competes with dilatancy \cite{wegner2014effects, hosseinpoor2021rheo}, a concept first introduced by Reynolds in the late 19\textsuperscript{th} century \cite{reynolds1885lvii}, which describes the volume expansion of material under shear due to particle interaction and the resulting relative motion. Consequently, sheared sticks experience two opposite effects: particle alignment, which causes a localized depression in the shear zone as particles pack more densely, and dilatancy, which leads to localized heaping due to the materials' volume expansion \cite{wegner2014effects,rahim2024alignment}.

When non-spherical particles are sheared in a split-bottom shear cell, they exhibit heap formation driven by dilatancy and secondary flows \cite{dsouza2021dilatancy, dsouza2017secondary, krishnaraj2016dilation}. Wortel et al. were the first to observe this phenomenon for a granulate of sticks, attributing the heap formation to the misalignment of the particles with the flow direction \cite{wortel2015heaping}. 

Both prolate and oblate ellipsoidal grains show this behavior, with the heap surface rising approximately 30\% higher than the initial filling, causing a depression of the surface above the shear band, a phenomenon not seen with spheres \cite{fischer2016heaping, wortel2015heaping, fischer2016heaping, rahim2024alignment}. Secondary flows have been extensively studied in Couette cells and split-bottom shear cells filled by densely packed spheres \cite{dsouza2021dilatancy, dsouza2017secondary, krishnaraj2016dilation}, as well as non-spherical grains of aspect ratio 1 \cite{mohammadi2022secondary}.
These studies have primarily focused on the effects of fill height, shear rate, and particle shapes on the dilatancy and secondary flows of non-elongated granular materials. However, the influence of friction and initial packing properties on the steady-state structure of sheared elongated particles in granular media remains largely unexplored.

Granular flow in cylindrical split-bottom shear cells has been studied to understand the influence of particle shape and material properties on secondary flows, orientational order, and dilatancy \cite{dijksman2010granular, fenistein2003wide, fischer2016heaping, wegner2014effects}. Building upon these results, recent studies have employed linear split-bottom shear cells (LSC) with periodic boundary conditions along the flow direction \cite{ries2007shear,dsouza2021dilatancy,roy2021drift, rahim2024alignment}. Despite the finite domain size, simulations employing periodic boundary conditions enable the study of infinitely extended systems by replicating the boundaries seamlessly, thus capturing the system's bulk behavior by suppressing effects due to the finite system size.

In the current paper, we report simulations of LSC. We investigate the impact of the particle shape, friction, and initial packing on the structure of granular flows. We analyze the steady-state packing density, surface profile, and the evolution of particle alignment within the shear band. The interplay between aspect ratio (AR) and friction ($\mu$) is examined to understand their influence on the steady-state surface profile of sheared sticks.

\section{Numerical methods and setup}
\subsection{Numerical methods}

We perform Discrete Element Method (DEM) simulations using the open-source code MercuryDPM \cite{weinhart2020fast,thornton2023recent}, which includes the multisphere algorithm \cite{ostanin2024rigid}, and advanced force approximations \cite{bagheri2024approximate,willett2000capillary, bagheri2025discrete}, making it a versatile platform for DEM simulations.


\subsection{System geometry}
\autoref{fig:LSC}(a) sketches the system geometry:
\begin{figure}[htbp]
\centering
\begin{subfigure}{0.99\columnwidth}
    \includegraphics[trim={0cm 0cm 0cm 0cm},clip,width=\linewidth]{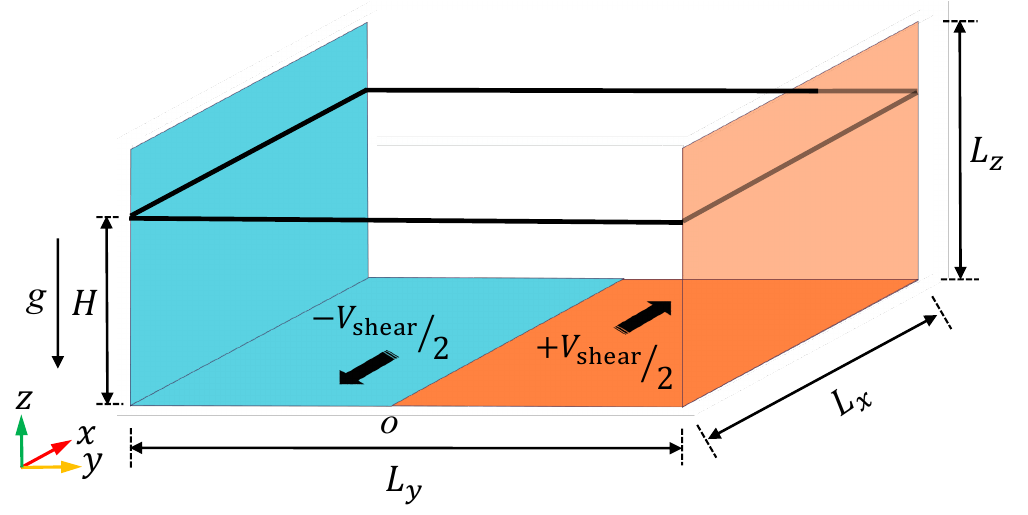}
    \subcaption{}
\end{subfigure}

\begin{subfigure}{0.9\columnwidth}
    \includegraphics[trim={0cm 0cm 0cm 0cm},clip,width=\linewidth]{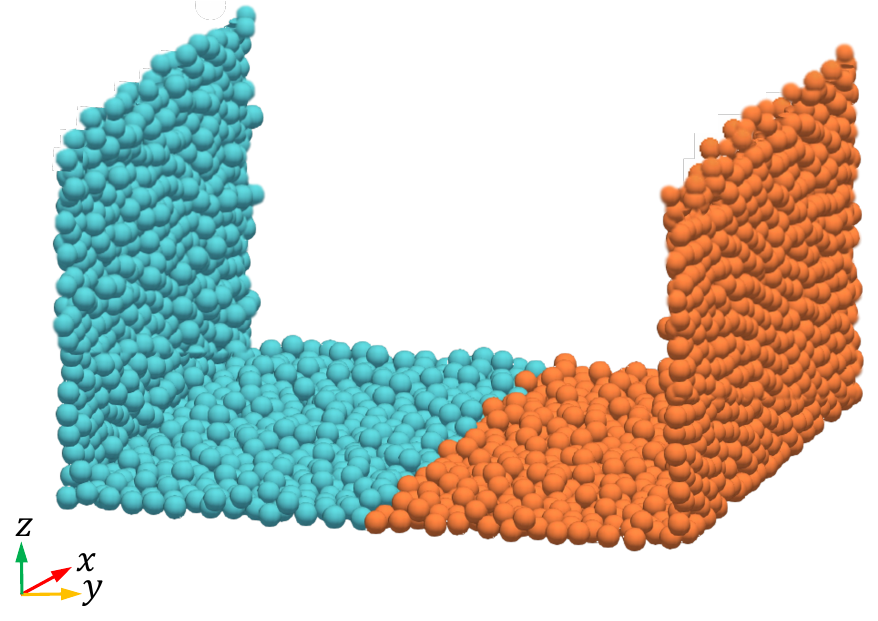}
    \subcaption{}
\end{subfigure}
\caption{(a) Sketch of the linear split-bottom shear cell (LSC): $y=0$ marks the split position, the shear velocity is $V_\text{shear}$, and $g$ marks the direction of gravity. (b) Spheres cover the walls to avoid slip.}
    \label{fig:LSC}
\end{figure}
The linear split-bottom shear cell consists of two L-shaped profiles. The position of the split, $y=0$, defines the location where the profiles move relative to each other in the $x$-direction at rate $V_\text{shear}$. Gravity $g$ acts in negative $z$-direction. The system parameters are inspired by the circular split-bottom experiment by Fischer et al. \cite{fischer2016heaping}. The shear velocity $V_\text{shear}=0.038\,\text{m/s}$ corresponds to 3 rpm, for $R_\text{disk}=118\,\text{mm}$ used in the experiment. The shear cell is periodic in $x$-direction. To avoid slip at the boundaries, the walls are modeled by spheres of diameter $d_{p} =8.55\,\text{mm}$, see \autoref{fig:LSC}(b). The size of the shear cell is $(L_{x}, L_{y}, L_{z}) = (25, 25, 20)\,d_{p}$, the filling height is $H$. When sheared, a shear band emerges from the split position, extending in $y$ and $z$ directions. In the following, we will present $y-z$-profiles where the data are averaged over the homogeneous $x$-direction.

\subsection{Particle shape}
\begin{figure}[htbp]
    \centering
    \includegraphics[trim={0cm 0cm 0cm 0cm},clip,width=0.99\columnwidth]{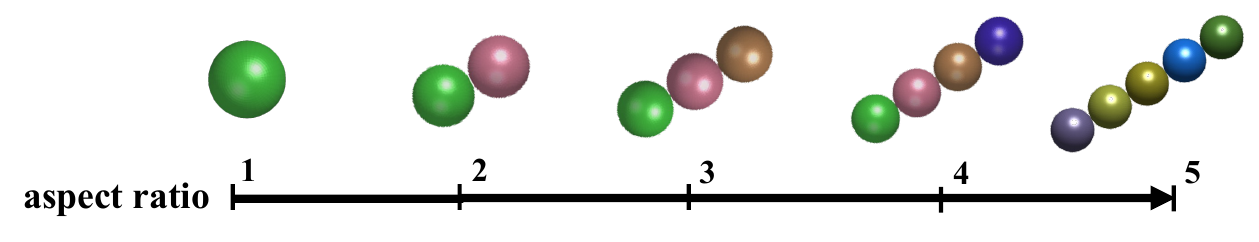}
    \caption{Stick-shaped particles used in the simulations. The number of spheres per stick defines the aspect ratio.}
    \label{fig:particle_shape}
\end{figure}

We consider elongated particles (sticks) formed by linearly arranged spheres as shown in \autoref{fig:particle_shape}. 
The number of spheres per stick defines the particle's aspect ratio (AR). The representation of non-spherical particles by connected spheres (multisphere approach \cite{Buchholtz:1993}) has the advantage that numerical contact detection simplifies considerably as it reduces to the contact of spheres \cite{abou2004three,kodam2009force,cabiscol2018calibration}. Spheres belonging to the same multisphere particle are identical. To study the influence of the aspect ratio, we keep the total volume of the packings, that is, the total mass of the granulate, invariant, independent of the AR. 
For AR=1, the granulate consists of $N=4,500$ spheres with radii chosen from an equidistribution $[0.8\,\left< d \right>, 1.2\,\left< d \right>]$ with mean $\left< d \right>=7.6\, \text{mm}$. To generate a granulate of sticks with $\text{AR}> 1$, we use the same set of random values, scale them by $\text{AR}^{1/3}$ and form $N=4,500$ sticks as sketched in \autoref{fig:particle_shape} each consisting of AR spheres. Consequently, the volume of the particles is independent of AR. The chosen number of particles, $N$, and their size distribution agree with the experiments reported in \cite{ries2007shear}.  
\subsection{Contact model and material parameters}
The visco-elastic Hertz-Mindlin contact model \cite{BSHP:1996, mindlin1949compliance} is used to describe the force between spheres in contact. The normal component reads \cite{BSHP:1996}
\begin{equation}
    \vec{F}_n = \min\left(0, -k\xi^{3/2} - \frac{3}{2}A_n k\sqrt{\xi}\dot{\xi}\right) \vec{e}_n\,,
\end{equation}
where $\xi\equiv R_i + R_j - \left|\vec{r}_i-\vec{r}_j\right|$ is the compression of spheres $i$ and $j$ of radii $R_i$ and $R_j$ at positions $\vec{r}_i$ and $\vec{r}_j$, $\vec{e}_n \equiv (\vec{r}_i-\vec{r}_j)/\left|\vec{r}_i-\vec{r}_j\right|$ is the normal unit vector. The normal dissipative parameter $A_n = 6\times 10^{-5}$s corresponds to the coefficient of restitution $0.4$ for a particle of radius $2.5\,\text{mm}$ and elastic modulus $ E = 10\,\text{MPa}$, impacting with velocity $2\,\text{m/s}$ \cite{muller2011collision}. 
The effective stiffness of the Hertzian contact model is
\begin{equation}
    k \equiv \frac{4}{3} \, E^* \, \sqrt{R^*} 
    \label{eq:Hertz_rho}
\end{equation}
with the effective radius $R^*$ and the effective elastic modulus
\begin{equation}
     E^* \equiv \left(\frac{1-\nu_i^2}{E_i} + \frac{1-\nu_j^2}{E_j}\right)^{-1}\,,
\end{equation}
where $E_{i}$ and  $\nu_{i}$ are the elastic modulus and the Poisson ratio of the material of particle $i$. 

For the tangential viscoelastic force, we assume the no-slip expression by Mindlin \cite{mindlin1949compliance} for the elastic part and Parteli and P\"oschel \cite{parteli2016particle} for the tangential dissipative constant $A_t \approx 2 A_n E^*$. The Coulomb criterion limits the force:
\begin{equation}
    \vec{F}_t = -\min \left[ \mu\left|\vec{F}_n\right|,  \int 8 G^{*}\sqrt{R^* \xi} \,ds 
    + A_t  \sqrt{R^* \xi} v_t \right] \vec{e}_t\,,
\end{equation}
with the friction coefficient, $\mu$, and the effective shear modulus 
\begin{equation}
    G^*=\left(\frac{2-\nu_i}{G_i} + \frac{2-\nu_j}{G_j}\right)^{-1},
\end{equation}
 which for identical materials simplifies to 
 \begin{equation}
     G^*=\frac{G}{2\left(2-\nu\right)}\,.
 \end{equation} 
 The integral is performed over the displacement of the colliding particles at the point of contact for the total duration of the contact \cite{parteli2016particle}.
The material parameters corresponding to wooden pegs are given in \autoref{tab:material_parameters} \cite{fischer2016heaping}.
\begin{table}[htb]
\caption{DEM simulation parameters}
\label{tab:material_parameters}
\begin{center}
\begin{tabular}{l@{\quad}l@{\quad}ll}
\hline
variable & unit& value\\
\hline
elastic modulus ($E$)  & MPa & 10\\
sliding friction coefficient ($\mu$)  & - & 0.01-0.8\\
Poisson's ratio ($\nu$)  & -& 0.35\\
particle density ($\rho$) & kg/m$^3$& 850\\
\hline                
\end{tabular}
\end{center}
\end{table}
For acceptable computer time, Young's modulus of the wooden pegs is reduced in the simulations. With the chosen value, the maximum achieved particle compression is less than 2\% of the particle diameter, which is within the acceptable limits for DEM simulations \cite{luding2008introduction, thornton2000numerical, cundall1979discrete}.

\subsection{Initial conditions}
The particles are filled into the LSC and relaxed until the kinetic-to-potential energy ratio falls below $10^{\text{-}4}$. This potential energy is the elastic potential energy stored in deformations of the particles in contact. Subsequently, we study the surface profile within the localized shear band region (around $y=0$) for the shear cell in its steady state, which is assumed to be independent of the initial conditions. To validate this assumption, we simulate the system's relaxation towards its steady state when starting from extreme initial conditions, termed IC-1 and IC-2. 

\begin{figure}[htbp]
\centering
    \includegraphics[trim={0cm 0cm 0.9cm 0cm},clip,width=0.98\columnwidth]{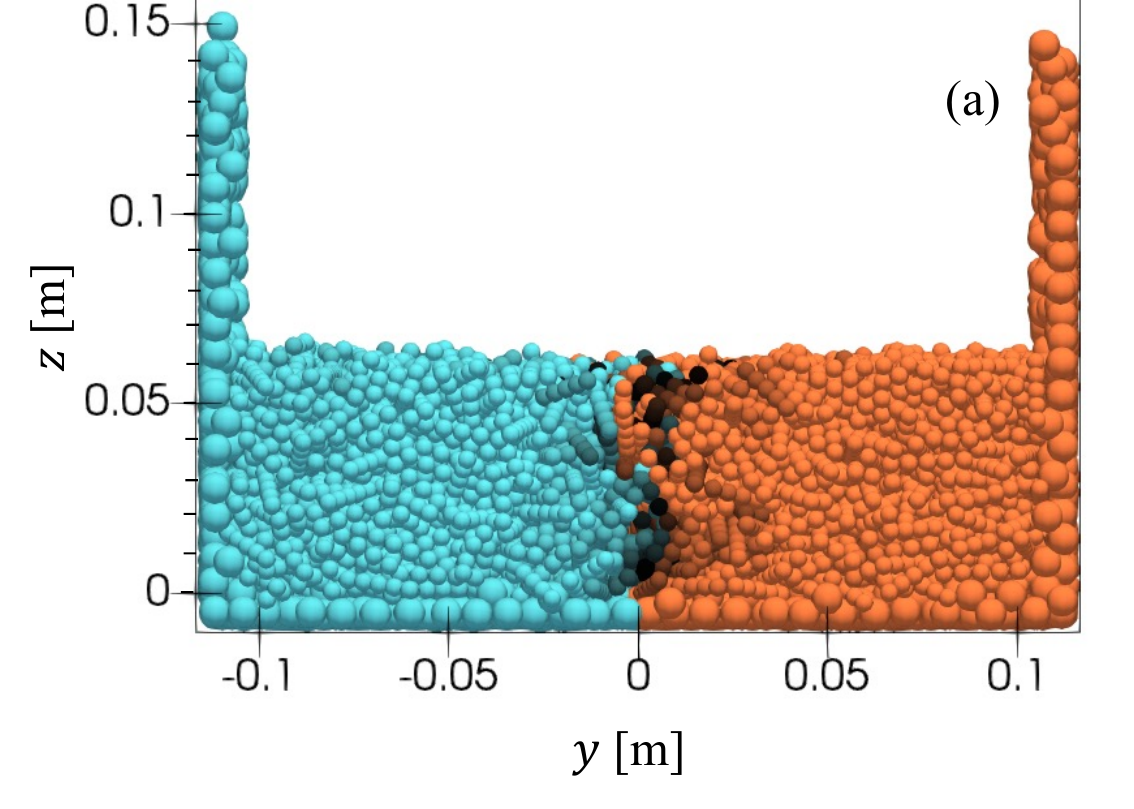}
    \includegraphics[trim={0cm 0cm 0cm 0cm},clip,width=0.98\columnwidth]{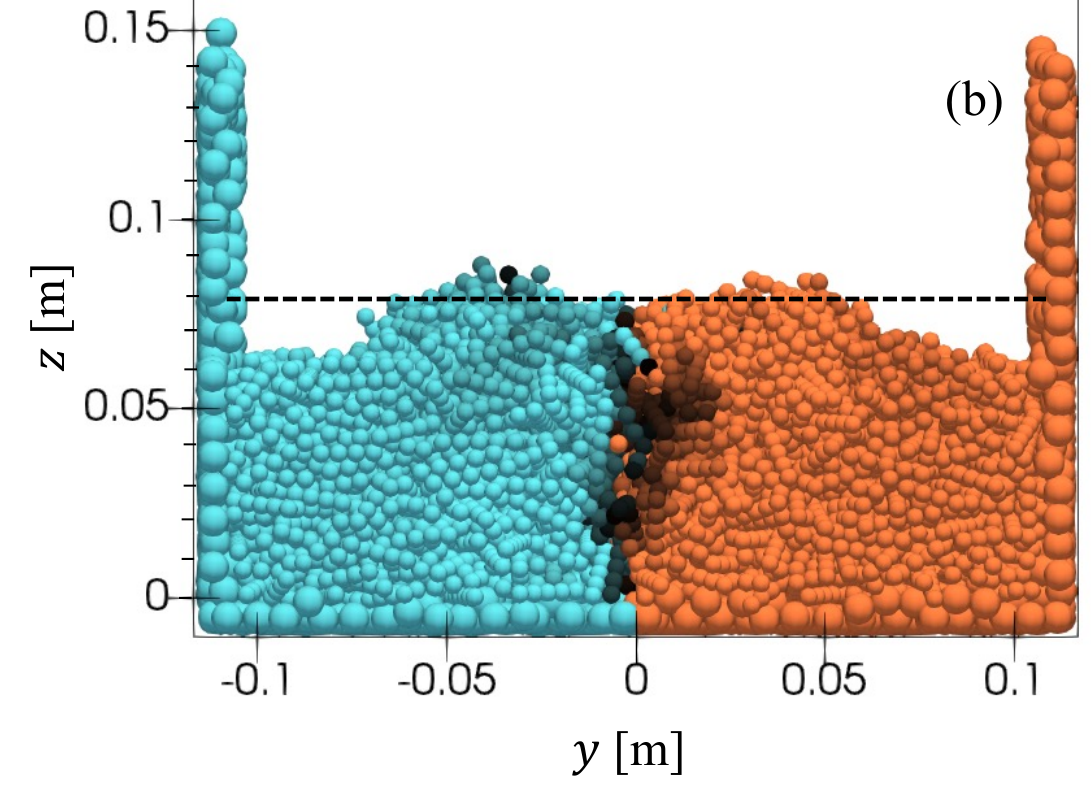}
    \hspace*{1cm} 
    \includegraphics[trim={0cm 0cm 0cm 0cm},clip,width=0.80\columnwidth]{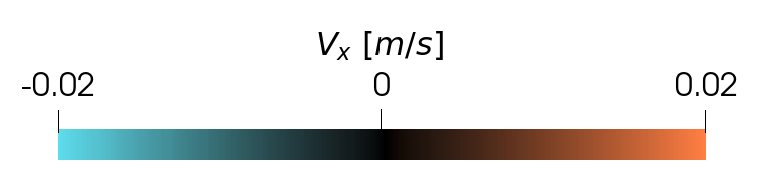}
    \caption{Snapshots of LSC simulations in the steady state, starting from IC-1 for sticks with AR = 5. Low friction (a) $\mu = 0.01$ results in a nearly flat surface. High friction (b) $\mu = 0.8$ results in a depression of the surface above the shear band region. The dashed line shows the packing height in the shear band region.}
    \label{fig:LSC-IC1}
\end{figure}

In IC-1, particles were deposited under gravity into the LSC with the lowest friction value $\mu = 0.01$ used in the simulations below, resulting in a uniform, dense initial packing. After the deposition, the friction coefficient is adjusted to its specified value in the range $0.01 \leq \mu \leq 0.8$ before the shear starts. \autoref{fig:LSC-IC1} shows the particle system for starting from IC-1 with $\mu = 0.01$ and $\mu = 0.8$, respectively, and $\text{AR} = 5$. Low friction leads to a nearly flat surface profile in the region above the shear band, whereas higher friction results in a depression at the center.

\begin{figure}[htbp]
    \includegraphics[trim={0cm 0cm 0cm 0cm},clip,width=0.98\linewidth]{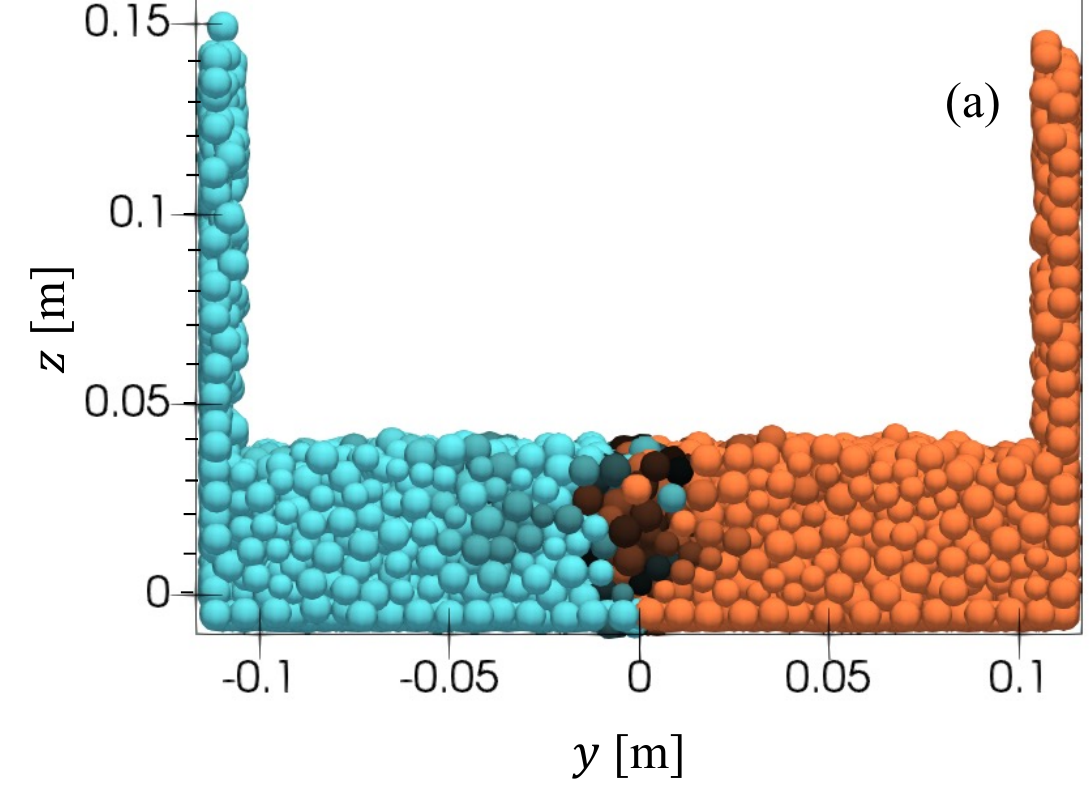}
    
    \includegraphics[trim={0cm 0cm 0cm 0cm},clip,width=0.98\linewidth]{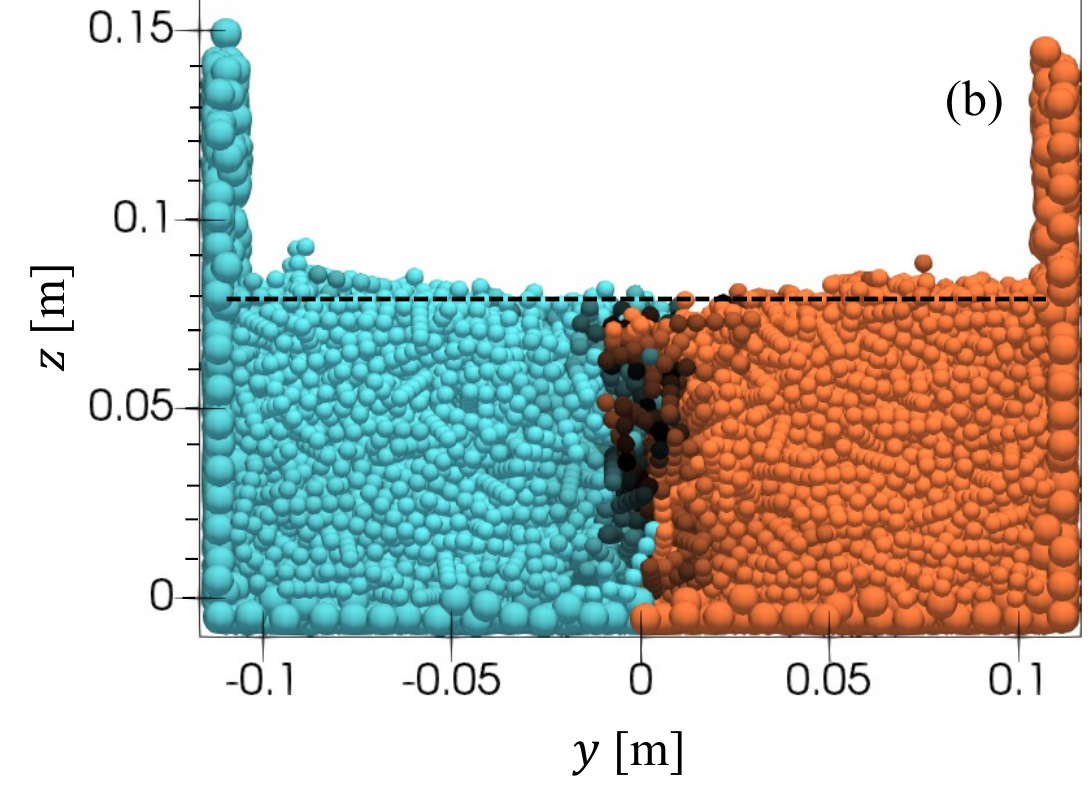}

    \hspace*{1.3cm} 
    \includegraphics[trim={0cm 0cm 0cm 0cm},clip,width=0.8\linewidth]{media/1_Geometry/colorbar_1.png}
    \caption{Snapshots of LSC simulations in the steady state, starting from IC-2 for particles with (a) AR=1 (spheres) and (b) AR=5. In both cases, $\mu = 0.8$. For spheres, the steady state reveals a flat surface above the shear band region; for sticks, we see a depression. The dashed line shows the packing height in the shear band region.}
    \label{fig:LSC-IC2}
\end{figure}

In IC-2, the particles are deposited using the same friction value, $\mu$, as in the shear process, $0.01 \leq \mu \leq 0.8$. Here, in contrast to IC-1, the initial filling height and, thus, the packing density for a given AR depend on $\mu$. Spherical particles ($\text{AR} = 1$) shown in \autoref{fig:LSC-IC2}(a) result in a flat surface profile, whereas the sticks ($\text{AR} = 5$) in \autoref{fig:LSC-IC2}(b) lead to a depression at the center of the shear band. 

\autoref{fig:LSC-IC1}(b) and \autoref{fig:LSC-IC2}(b) show snapshots of the system in the steady state for the same set of parameters, $\mu = 0.8$ and AR=5, but for different initial conditions, IC-1 and IC-2. We see that the surface height in the shear band region, indicated by dashed lines is nearly the same, and the depression is visible in both figures, thus, the simulation results \textit{for the shear band region} are independent of the initial conditions. Outside the shear band region, the simulation results differ for IC-1 and IC-2. This is due to the extremely slow flow velocity outside the shear band, leading to very long relaxation times. Asymptotically, we expect that the system approaches the same steady state, independent of the initial conditions. This, however, exceeds the limits of numerical simulations and possibly even the limits of experiments. We conclude that our simulations reveal the steady-state flow in the shear-band region but perhaps only a transient state outside the shear-band region. In the following, we will present results regarding the shear-band region, thus, the transient nature of the system outside this region is irrelevant to our results. 

For the results below, we use IC-2 initial conditions, in agreement with the experiments \cite{liu2017experimental}.

\subsection{Coarse-graining of stationary fields}
\label{sec:cg}
To extract the macroscopic fields, we used a coarse-graining tool MercuryCG \cite{weinhart2016influence} to obtain the macro-parameters through precise calculation of sphere overlap volumes and mesh elements, as outlined by Strobl et al. \cite{strobl2016exact}. We simulated the process for 300\,s real-time. The time-dependent shear displacement is $\lambda = V_{x}\,t$. For all sets of parameters used, we found that the kinetic energy and average contact number approach their steady-state values before $\lambda \approx 47 \, d_p$ (10 s). The fully developed stationary flow profile, including the shear band, was achieved in all cases before $\lambda \approx 280 \, d_p$ \cite{singh2024shear, ries2007shear}. The fields of strain rate, velocity, packing density, and stress were obtained from the particle trajectories by coarse-graining, based on the exact intersection of the spheres with spatial mesh elements \cite{strobl2016exact} (see \cite{ganthaler2024comparison} for a comparison of coarse-graining methods).

The fields are averaged over the (periodic) $x$-direction in the time interval $t\in (250, 300)\,\text{s}$ when the system adopted its stationary state. From these fields, we obtained the stationary quantities studied here. Of particular interest is the shear band region. It is defined as the region in the $y-z$ plane where the local strain rate at a height \( z \) exceeds $\dot{\gamma}_c(z) \equiv 0.8 \dot{\gamma}_{\text{max}}(z)$, where $\dot{\gamma}_{\text{max}}(z)$ is the maximum strain rate at height $z$.

\section{Packing density}
\subsection{Packing density for spherical particles}
\begin{figure}[htb]
\centering
\includegraphics[trim={0cm 0cm 0cm 0cm},clip,width=0.99\columnwidth]{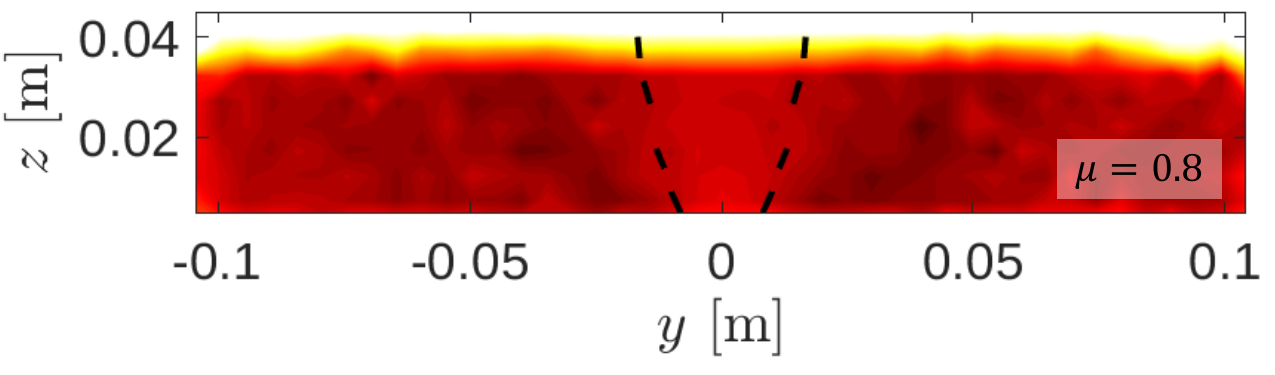}
    \hspace*{1.2cm} 
    \includegraphics[trim={0cm 0cm 0cm 0cm},clip,width=0.75\columnwidth]{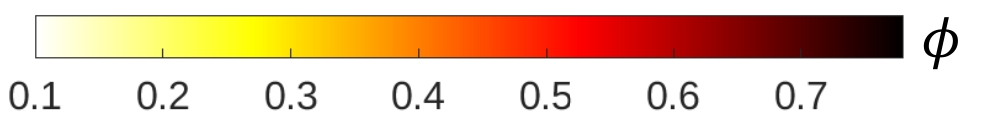}
\caption{Field of the packing density, $\phi(y,z)$, in the stationary state for $\mu = 0.8$ and $\text{AR} = 1$ (spheres). The dashed lines indicate the location of the shear band, as defined in Sec. \ref{sec:cg}.} 
\label{fig:PD_Sphere}
\end{figure}

\autoref{fig:PD_Sphere} shows the field of the packing density, $\phi(y,z)$, for spheres (particles with $\text{AR} = 1$) with $\mu = 0.8$.
The dashed lines indicate the shear band boundaries. The width of the shear band is determined by fitting the normalized velocity profile in the $x$ direction as a function of $y$ to an error function \cite{dijksman2010granular, ries2007shear, depken2006continuum}. The width is then determined from the error function fit.

We obtain a flat surface profile above the shear band region for all studied values of $\mu\in [0.01,0.8]$. In particular, the fields are invariant with respect to the initial conditions.

\subsection{Packing density of sticks}

\begin{figure}[htbp]
\centering
\begin{subfigure}{0.90\columnwidth}
    \includegraphics[trim={0cm 0cm 0cm 0cm},clip,width=\linewidth]{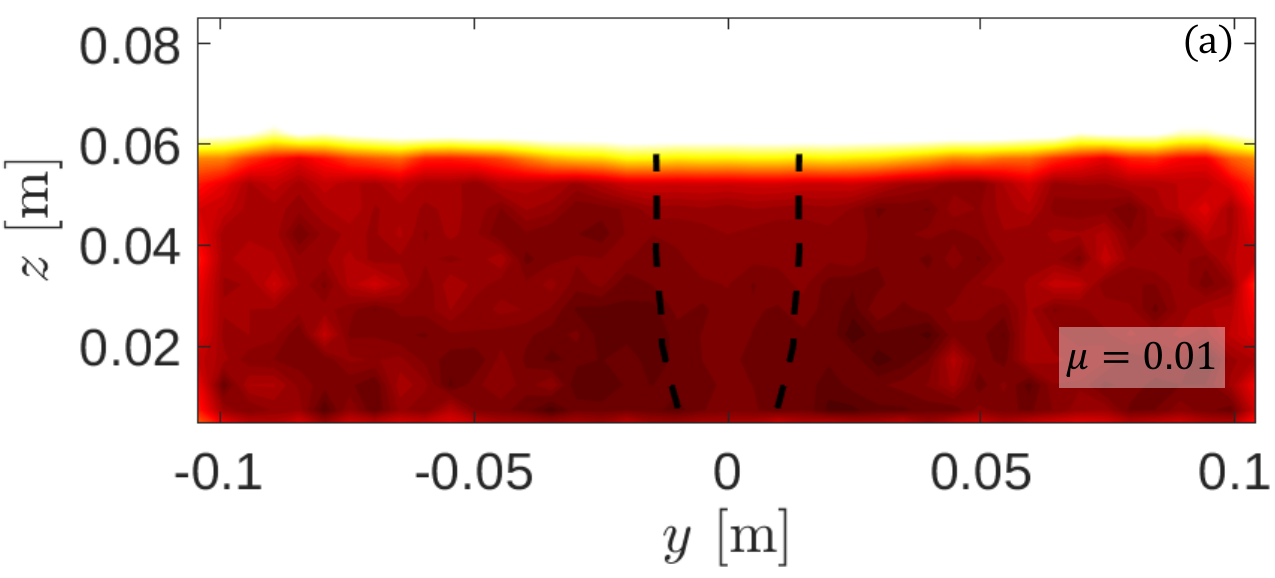}
\end{subfigure}
\hfill
\begin{subfigure}{0.90\columnwidth}
    \vspace{1ex}
    \includegraphics[trim={0cm 0cm 0cm 0cm},clip,width=\linewidth]{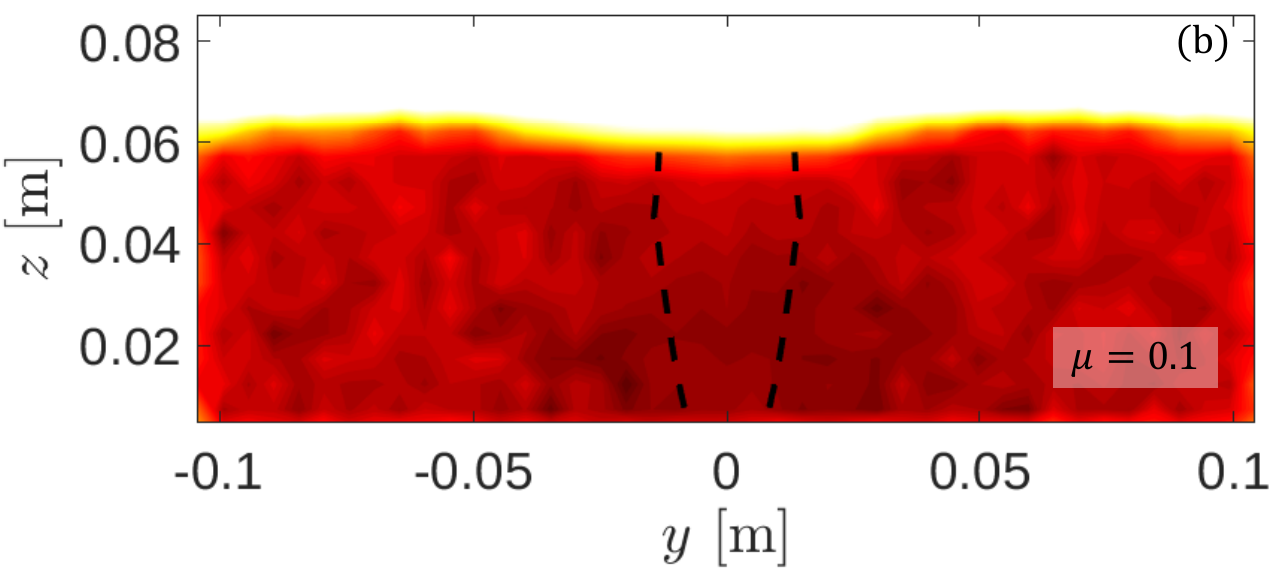}
    \\[1ex]
\end{subfigure}
\begin{subfigure}{0.90\columnwidth}
    \vspace{1ex}
    \includegraphics[trim={0cm 0cm 0cm 0cm},clip,width=\linewidth]{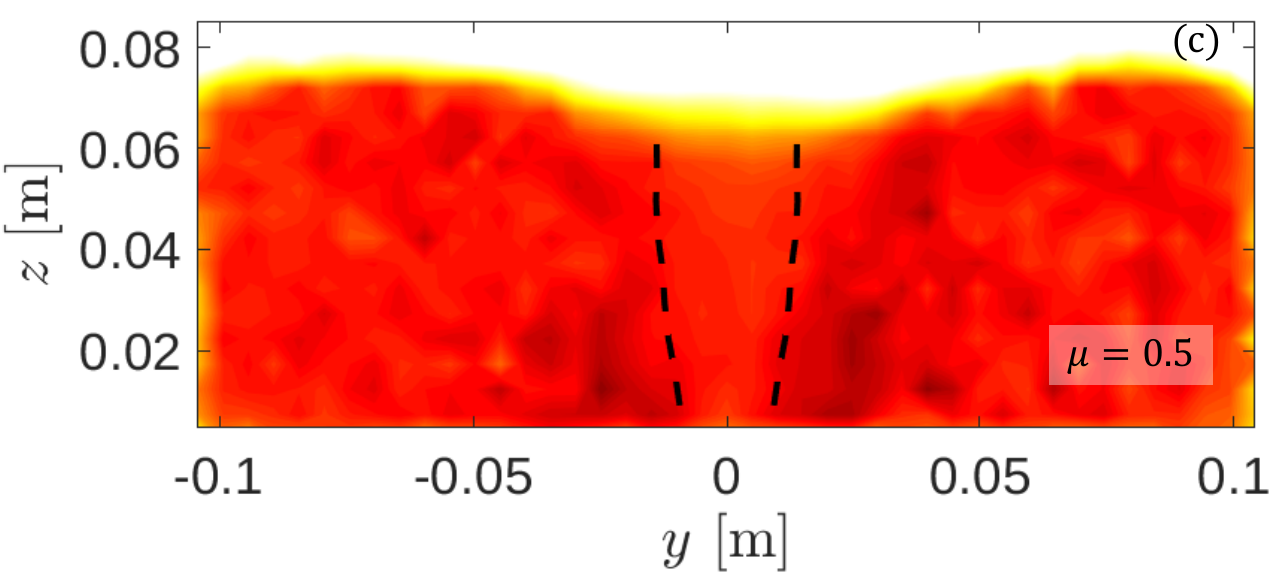}
\end{subfigure}
\hfill
\begin{subfigure}{0.90\columnwidth}
    \vspace{1ex}
    \includegraphics[trim={0cm 0cm 0cm 0cm},clip,width=\linewidth]{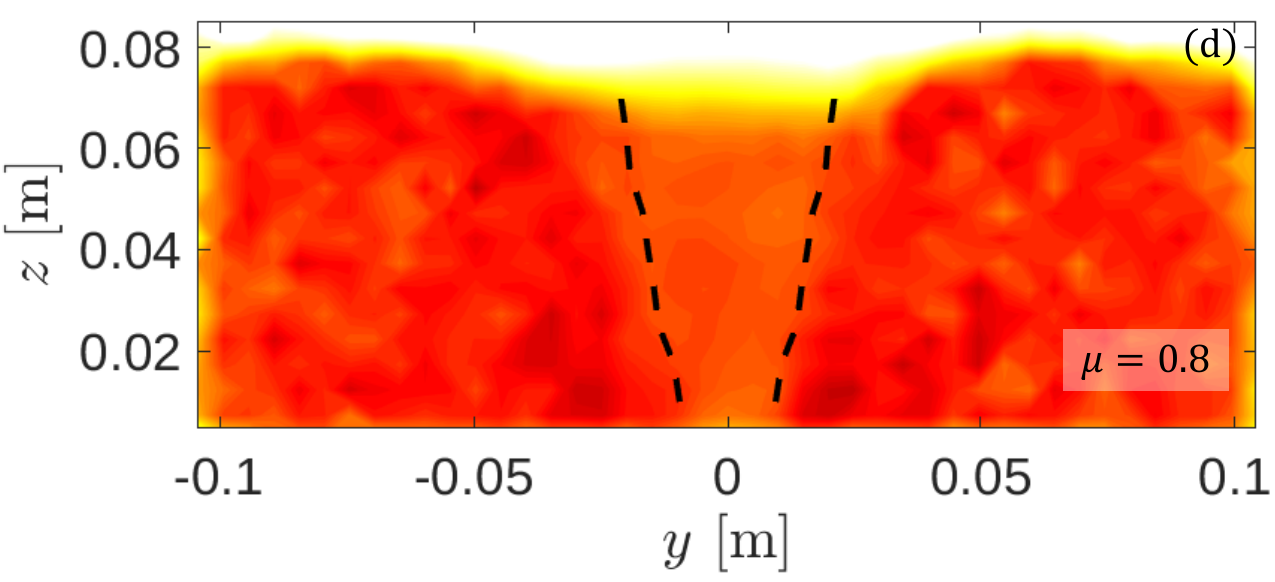}
\end{subfigure}

\hspace*{0.9cm} 
\begin{subfigure}{0.8\columnwidth}
    \includegraphics[trim={0cm 0cm 0cm 0cm},clip,width=\linewidth]{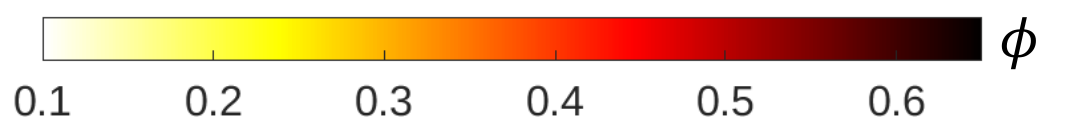}
\end{subfigure}

\caption{Field  of packing density in the stationary state for $\text{AR} = 5$ and different $\mu$ in the $yz$-plane, averaged over the $x$-direction. The dashed lines indicate the location of the shear band, as defined in Sec. \ref{sec:cg}}
\label{fig:PD_IC-2}
\end{figure}

\autoref{fig:PD_IC-2} 
shows the packing density field in the $yz$ plane averaged in the $x$-direction, for $\text{AR} = 5$. The dashed lines indicate the location of the shear band. The value of $\mu$ in the deposition process determines the initial filling height and packing density before shearing begins. At low friction ($\mu = 0.01$), the sticks can easily move and reorient within the shear band region, leading to their alignment along the shear direction and, thus, to a compact arrangement corresponding to denser packing and, eventually, the subtle surface depression, seen in \autoref{fig:PD_IC-2}(a-b).
The particle motion is hindered as $\mu$ increases. However, the initial loose packing provides sufficient space for the sticks to move and reorient within the shear band region. This alignment leads to compaction and depression formation at the shear band surface, \autoref{fig:PD_IC-2}(c-d). 

This indicates that two competing effects occur within the shear band: alignment and dilatancy. Alignment leads to increased density, while dilatancy causes the materials to expand, decreasing the density. For small $\mu$, the effects of particle alignment dominate over dilatancy, stabilizing the microstructure and resulting in a large packing density, as shown in \autoref{fig:PD_IC-2}(a-b). Conversely, dilatancy becomes more pronounced for large $\mu$, disrupting particle alignment and reducing the packing density, as shown in \autoref{fig:PD_IC-2}(c-d).

\subsection{Packing density in the shear band region}

Sticks exhibit depression, while spherical particles maintain a flat surface profile under shear. To understand this behavior, we study the local density as a function of the stress ratio, $\phi(\tau/p)$, in the shear band region, where 
\begin{equation}
    \tau \equiv \sqrt{\sigma_{xy}^2 + \sigma_{xz}^2}
\end{equation} 
is the local shear stress and 
\begin{equation}
    p \equiv \frac{1}{3} (\sigma_{xx} + \sigma_{yy} + \sigma_{zz})
\end{equation} 
is the local normal stress. $\sigma_{xx}$, $\sigma_{yy}$, and $\sigma_{zz}$ are the normal stresses in shear, lateral, and orthogonal directions, respectively, and $\sigma_{xy}$ and $\sigma_{xz}$ are the shear stresses in the shear plane. 

\begin{figure}[htbp]
    \centering
    \includegraphics[trim={0cm 0cm 0cm 0cm},clip,width=0.90\columnwidth]{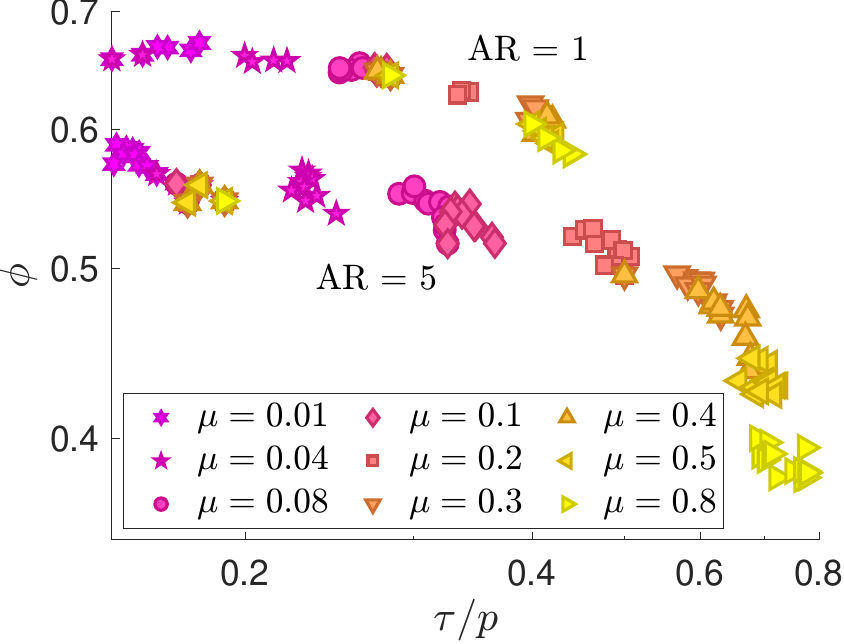}
    \caption{Packing density in the shear band as a function of the stress ratio, $\phi(\tau/p)$ for spheres and sticks and several values of $\mu$. For sticks ($\text{AR}=5$), the decay of the function is more pronounced than for spheres.}
    \label{fig:PD_crit_state}
\end{figure}

We find that the local packing density decays with increasing $\tau/p$ for both spheres and sticks, see \autoref{fig:PD_crit_state}. This dependence is more pronounced for sticks with $\text{AR} \ne 1$ than for spheres. 

Despite higher packing density, spherical particles maintain a flat surface profile due to their symmetrical shape, which allows them to easily rotate, translate, and slide past each other under shear, stabilizing into a new arrangement. Conversely, for sticks the packing density depends sensitively on friction, which affects the surface profile along the $y$-direction.

\section{Particle alignment in the shear band region}

The angle 
\begin{equation}\label{dotprod}
   \theta_{x} \equiv \frac{\pi}2-\left|\frac{\pi}2- \arccos \left(\frac{\vec{P_{v}} \cdot \vec{X}}{\left|\vec{P_{v}}\right| \left|\vec{X}\right|}\right)\right|
\end{equation}
quantifies the deviation of stick's principal vector, $\vec{P_{v}}$, from the shear direction, $\vec{X}$. The limits, $\theta_{x}=0$ and $\theta_{x}=\pi/2$, stand for perfect alignment and misalignment, respectively. We consider the average over particles in the shear band region, $\left<\theta_{x}\right>$.

\begin{figure}[htbp]
\centering
        \includegraphics[trim={0cm 0cm 0cm 0cm},clip,width=0.90\columnwidth]{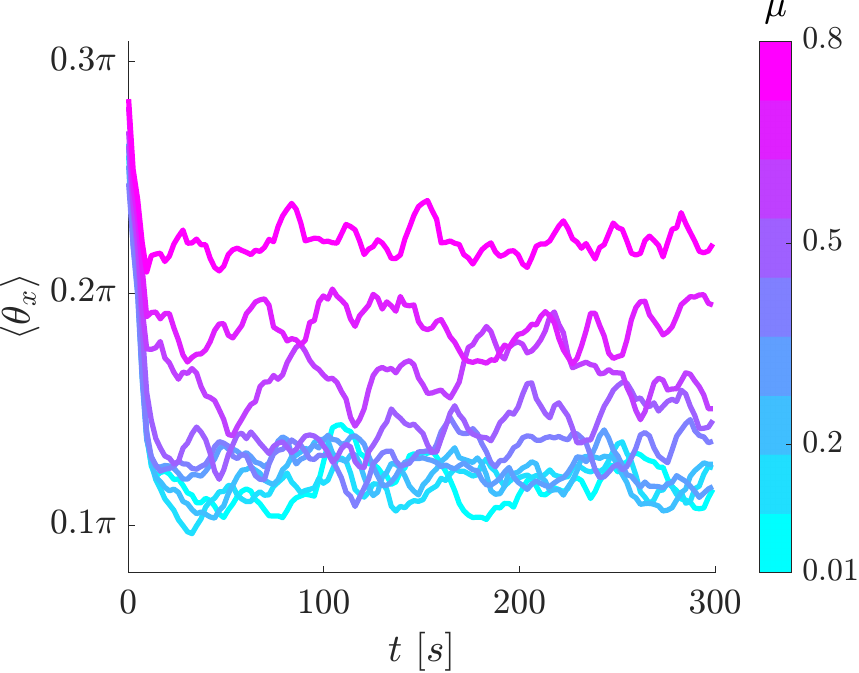}
        \hspace*{1cm}(a)

        \includegraphics[trim={0cm 0cm 0cm 0cm},clip,width=0.9\columnwidth]{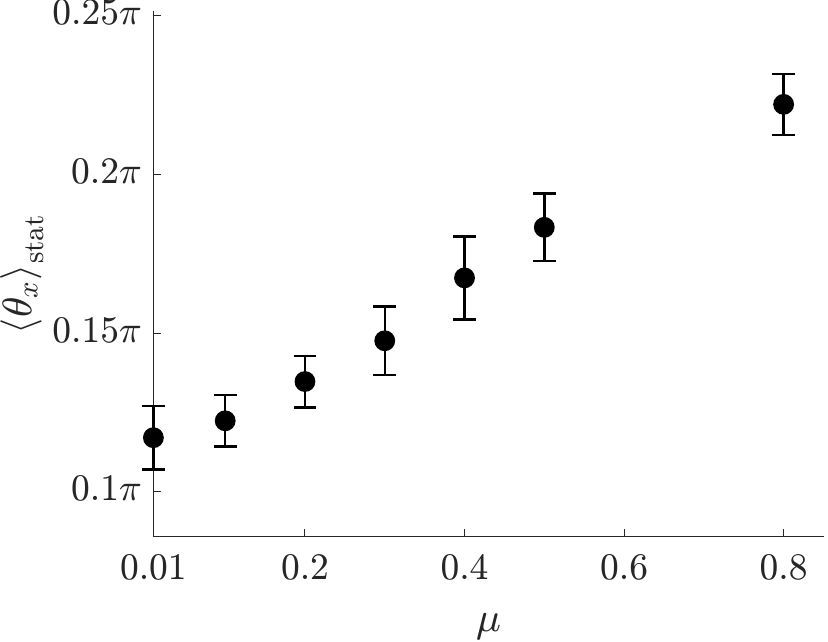}
        \hspace*{1.5cm}(b)

        \caption{(a) Alignment of particles in the shear zone region as a function of time, $\left<\theta_{x}\right> (t)$ for sticks with $\text{AR} = 5$ and different values of $\mu$. (b) Stationary values $\left< \theta_x \right>_{\mathrm{stat}}$ as a function of $\mu$.}
    \label{fig:Alignment_IC_all}
\end{figure}

In the simulation, we see a quick relaxation of $\left<\theta_{x}\right>$ as a function of time, starting from the initial value $\left<\theta_{x}\right>^{(0)}\approx \pi/4$ to a stationary state value which depends on $\mu$, see \autoref{fig:Alignment_IC_all}(a). 
For small $\mu$, the alignment appears more pronounced than for large $\mu$. Since the observed depression in the shear band region (see \autoref{fig:PD_IC-2}) is caused by the dense compaction due to alignment, we should expect a more significant depression for small $\mu$. This is not the case. Instead, the depression is more substantial for larger $\mu$, which can be understood from the width of the shear band: for small $\mu$, the wider shear band leads to a more uniform and gradual surface descent over a larger area. This can be seen in \autoref{fig:PD_IC-2}(a), where at $\mu = 0.01$, density is large with a shallow and wide depression, in contrast to $\mu = 0.8$ in \autoref{fig:PD_IC-2}(d), where the packing is less dense with a pronounced depression in the shear band region. 

\autoref{fig:Alignment_IC_all}(b) shows the stationary value of $\left< \theta_x \right>$ as a function of $\mu$, obtained from the simulation data in the interval $t\in [150, 300]\,\text{s}$. For $\mu = 0.01$, $\langle \theta_x \rangle_{\mathrm{stat}}$ ($\approx 0.12\pi$) is 47\% less than for $\mu = 0.8$ ($\approx 0.23\pi$). This increase in $\langle \theta_x \rangle_{\mathrm{stat}}$ with $\mu$ shows a less pronounced alignment of the sticks with the shear direction.

\section{Area of the flow profile}
\label{sec:heap_area}
\subsection{Relaxation}

Consider the total area of the shear band profile,
\begin{equation}
A= \int\limits_y z_{\text{max}}(y) \, dy
\end{equation}
as a function of time. The integral is performed over the entire range of $y$, and $z_{\text{max}}(y)$ is the position of the uppermost particle at a specific location, $y$, averaged over the shear direction, $x$. The initial value, $A_0=A(0)$, is assumed after the initialization and corresponds to a homogeneous surface.

\begin{figure}[htbp]
\centering
\includegraphics[trim={0cm 0cm 0cm 0cm},clip,width=0.9\columnwidth]{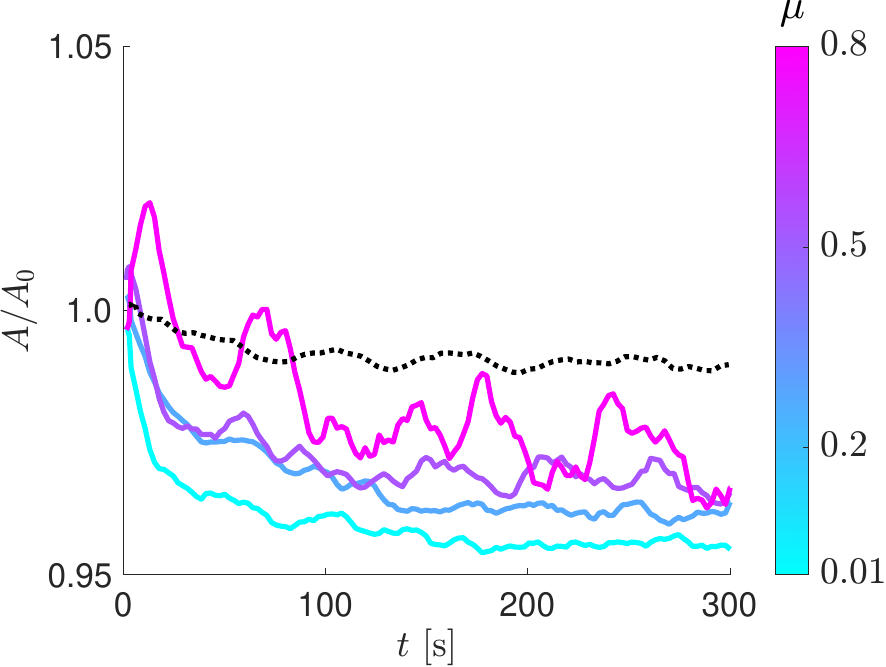}
\caption{$A/A_0$ as a function of time for sticks ($\text{AR} = 5$) with various $\mu$. The black dotted line shows the data for spheres and $\mu = 0.8$.}
\label{fig:heap_area} 
\end{figure}

\autoref{fig:heap_area}(b) shows $A/A_0$ over time for sticks ($\text{AR} = 5$, $\mu\in\{0.01,0.8\}$) and spheres  with $\mu = 0.8$ (dashed line). For all particles, $A/A_0$ decays and relaxes to a steady state value, indicating compaction due to particle alignment. 

\subsection{Effect of \texorpdfstring{$\mu$}{mu} and AR on the flow profile}

\begin{figure}[htbp]
\centering
\begin{subfigure}{0.9\columnwidth}
    \includegraphics[trim={0cm 0cm 2.26cm 0cm},clip,width=\columnwidth]{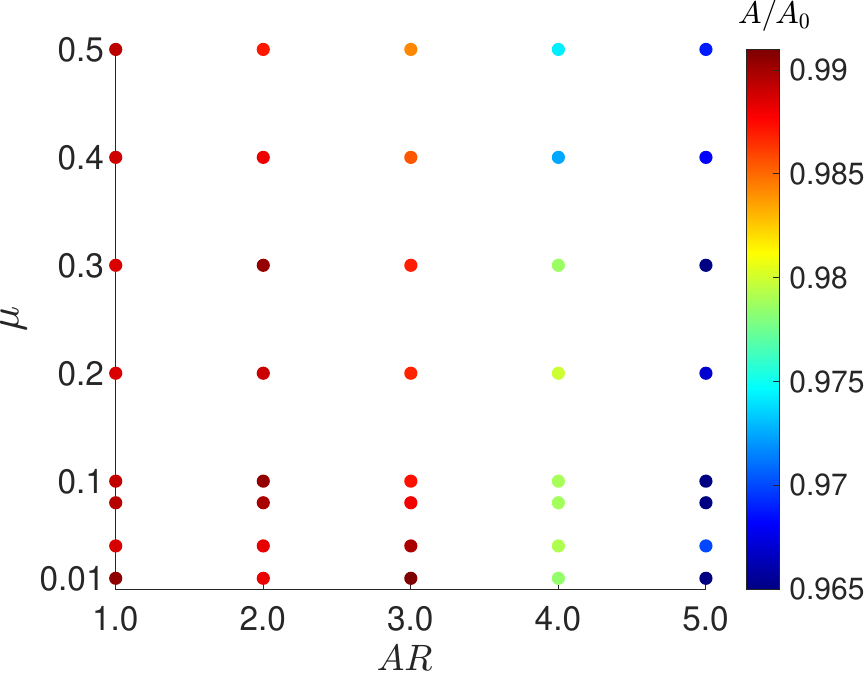}
    \hspace*{1.3cm} (a)
\end{subfigure}
\begin{subfigure}{0.90\columnwidth}
    \includegraphics[trim={0cm 0cm 1.7cm 0cm},clip,width=\columnwidth]{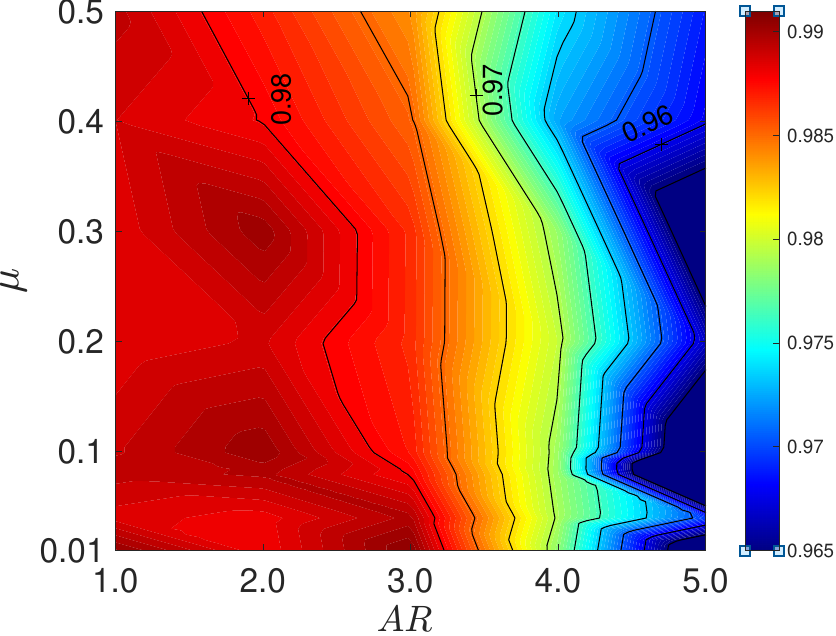}
    \hspace*{1.3cm} (b)
\end{subfigure}

\hspace*{0.5cm}
\begin{subfigure}{0.8\columnwidth}
    \includegraphics[trim={0cm 0cm 0cm 0cm},clip,width=\columnwidth]{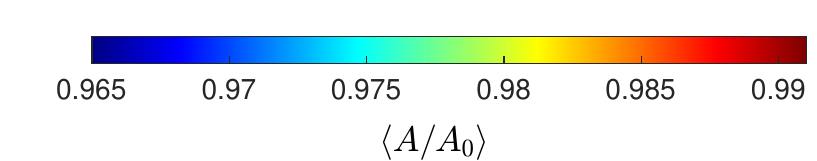}
\end{subfigure}

\caption{(a) Steady-state area ratio ${\langle A/A_0 \rangle}$ as shown in the color bar, plotted against AR and $\mu$. (b) Contour plot created from the data in (a) with color representing the stationary values of $\langle A/A_0 \rangle$.}
\label{fig:Phase_diagram} 
\end{figure}

\autoref{fig:Phase_diagram} shows the steady state value of $\left< A/A_0 \right>$ as a function of the $\text{AR}\in [1, 5]$ and $\mu \in [0.01, 0.5]$. The steady-state values were obtained from the simulation data in the interval $t\in [150, 300]\,\text{s}$ when the system converged to the steady state.

The color of the data points in \autoref{fig:Phase_diagram}(a) characterize their $\left< A/A_0 \right>$ value. 

Using all points in \autoref{fig:Phase_diagram}(a), we interpolate ${\langle A/A_0 \rangle}$, and obtain the contour plot as shown in \autoref{fig:Phase_diagram}(b).
Dark red regions at lower $\text{AR}$ (1 to 3) indicate $\langle A/A_0 \rangle$ values close to 1, suggesting minimal area depression due to reduced particle alignment. For $\text{AR}$ (4, 5), the color changes from yellow to blue, demonstrating a pronounced depression caused by increased particle alignment \cite{rahim2024alignment}. This indicates that the particle’s \(\text{AR}\) is the main factor influencing depression formation on the shear band surface.
\autoref{fig:Phase_diagram}(a) and (b) are based on 40 simulations with $\text{AR} \in [1, 5]$ and $\mu$ varying from 0.01 to 0.5.

We simulated only integer AR values, applying interpolation for better visualization. Previous studies \cite{rahim2024alignment} show that alignment-driven depression formation is approximately similar for elongated particles made of overlapping spheres (integer AR) and non-overlapping spheres (non-integer AR). Thus, intermediate AR is expected to exhibit similar behavior. However, the number of simulations is limited due to high computational costs.


\section{Conclusions}

The interplay between particle friction and AR determines the steady-state structure of sheared granular materials with non-spherical particles, where shape-induced alignment and friction-driven dilatancy have opposing effects on packing density and surface morphology.

At low friction, shape-induced particle alignment dominates, resulting in denser packing and a wider uniform surface depression than at higher friction. Conversely, at higher friction, dilatancy becomes more pronounced, disrupting particle alignment and reducing packing density, leading to localized surface depressions.

The ratio $A/A_0$ correlates with the dominant phenomena within the shear band. $A/A_0 < 1$ correspond to alignment-driven compaction while $A/A_0 > 1$ signifies dilatancy. For elongated particles, ${A}/{A_0}$ remains below $1$ for all $\mu$, indicating alignment-induced compaction. A more pronounced alignment at low friction leads to a denser packing and lower ${A}/{A_0}$. At high friction, dilatancy expands the material, increasing ${A}/{A_0}$.

For small AR particles, surface profile changes are minimal compared to higher AR particles due to their symmetric shapes, which resist alignment or dilation under shear. In contrast, sticks ($AR>3$) exhibit pronounced alignment-driven compaction and depression formation influenced by friction. This behavior indicates that AR is the primary determinant of the surface profile evolution under shear.




\section{Acknowledgement}
We thank Holger Götz and Thomas Weinhart for stimulating discussion. Financial support through
the DFG grant PO472/40-1 is acknowledged.

\bibliography{references}
\end{document}